Automated Dose-Based Anatomic Region Classification of Radiotherapy Treatment for Big Data Applications

Justin Hink[1], Yasin Abdulkadir[1], Jack Neylon[1], Jim Lamb[1]

[1]Department of Radiation Oncology, David Geffen School of Medicine at UCLA, University of California, Los Angeles, Los Angeles, California, USA

**Abstract**

Curation is a significant barrier to using 'big data' radiotherapy planning databases of 100,000+ patients. Anatomic site stratification is essential for downstream analyses, but current methods rely on inconsistent plan labels or target nomenclature, which is unreliable for multi-institutional data. We developed software to automate labeling by inferring anatomic regions directly from dose-volume overlap with deep-learning segmentations, eliminating metadata reliance. The software processes DICOM files in bulk, utilizing deep learning to segment 118 structures (organs, glands, and bones) categorized into six regions: Cranial, Head and Neck, Pelvis, Abdomen, Thorax, Extremity. The 85% and 50% isodose lines are converted to structures to compute organ-specific dose-overlap metrics. Plans are assigned ranked regional labels based on these intersections. The algorithm was refined using 109 expert-labeled cases and validated on 100 consecutive clinical plans. On the 100-plan test dataset, the algorithm achieved 91% Exact Accuracy (matching all expert labels and order), 94% Top-2 Accuracy (matching the top two expert regions regardless of order), and 95% Top-1 Accuracy (matching the primary expert label). The automated workflow demonstrated high accuracy and robustness. The 95% Top-1 Accuracy is particularly significant, as it enables reliable querying of plans based on the primary treatment site. Detailed analysis of the few mismatched cases showed most were treated areas at the border between anatomic regions and were ambiguous between these two regions in a common-sense interpretation. This algorithm provides a scalable, standardized solution for curating the large, multi-institutional datasets required for 'big data' in radiotherapy and provides an important complement to text-based approaches.

## 1. Introduction

Data curation is an essential task in gaining insight from large clinical databases. This is true when the data is single-institutional, and even more so when the database is formed by joining data from multiple clinics[1]. Increasingly, radiotherapy planning data is being gathered to unlock the potential of "Big Data" for analyzing treatment outcomes and automating complex tasks, such as tumor contouring[2]. Public "Grand Challenges", such as the Brain Tumor Segmentation (BraTS) Challenge[3], have successfully driven innovation by providing highly curated, multi-institutional datasets. However, public radiotherapy datasets are scarce. Such datasets are generally curated from the internal databases of academic institutions. This presents a unique challenge: if a clinic intends to share a dataset for a specific disease site, it must retrospectively identify hundreds to thousands of such plans from within its internal database. This is rarely straightforward with real-world clinical databases often suffering from inconsistent nomenclature, shifting labeling protocols, and non-standardized metadata, making simple queries unreliable[4].

As demonstrated in previous work[5] involving the bulk collection of treatment plans from diverse Veteran's Affairs' systems, overcoming these raw data challenges necessitates radiotherapy plan curation. This systematic process of cleaning, standardizing, and verifying complex DICOM objects ensures the data can be reliably used for downstream research. A key first step in radiotherapy plan curation is determining what part of the body was targeted because this anatomical site classification is upstream of essentially all further analysis[6]. When performed manually, accurately labeling radiotherapy plan anatomic region is a demanding and time-consuming task that requires specialized knowledge[4]. Traditional manual labeling is prone to human error due to its repetitive nature, leading to inaccuracies stemming from subjectivity and fatigue[7]. Because specialized clinical knowledge is required, the task is not amenable to outsourcing to low-cost workers, as has been used effectively for natural images[8]. With individual institutions able to gather large clinical repositories containing up to 20,000 patient datasets or more, curating these plans manually is infeasible.

Significant strides in automating annotation tasks have been made in other fields, particularly Computer Vision and Natural Language Processing. Broadly speaking, three classes of approaches have been used: fully automated deep learning models trained using supervised learning[9,10], semi-supervised approaches where human experts correct machine-generated labels[11-13], and unsupervised algorithms that establish data-driven classes[14,15].

In the context of radiotherapy, the use of standardized nomenclature (such as AAPM TG-263[16]) has ameliorated anatomic region labeling, but purely lexicographical approaches remain incomplete solutions. While advanced Natural Language Processing (NLP) and Large Language Models (LLMs) can parse plan names and ROI labels to detect naming patterns[17], they are susceptible to ambiguity. For example, generic labels like "New_Structure" or "ROI_1" offer no semantic clues for an LLM to interpret[4]. More importantly, text labels describe intent, whereas the physical treatment delivery is defined by the image data. Relying on metadata risks misclassifying plans where the label does not accurately reflect the dose distribution geometry[18,19].

To address these limitations, this study proposes an image-based methodology that utilizes the planned dose, the CT simulation, and deep learning-based normal-organ autocontouring to quantitatively determine the target anatomic region. This workflow will determine the feasibility of automated labeling for big databases on the order of 100,000 patients.

## 2. Methods

The proposed methodology is implemented as a three-stage automated workflow: (1) autosegmentation of organs-at-risk (OARs), (2) dose-mask-OAR intersection analysis, and (3) anatomic region labeling based on computed overlaps. This workflow quantifies the physical intersection between the delivered dose and anatomy. This geometric dose-based criterion provides an objective and reproducible measure of treatment site that is independent of text metadata. The overall goal was to automatically classify each treatment plan into one or more anatomic regions (Pelvis, Abdomen, Thorax, Cranial, Head and Neck, Extremity) based on which organs are overlapped by the target dose-wash.

## 2.1. Organ Autosegmentation and Preprocessing
### 2.1.1. Data Preparation and Preprocessing

Required input data include CT slices, RTDose, RTStruct, and RTPlan files in DICOM format. While the RTStruct is not utilized for its contours (as this workflow relies on deep-learning autosegmentation), it defines an essential connection between the RTPlan and the CT image series. Due to the inherent complexity of DICOM associations[5,20] and the non-trivial linking required between these objects, we utilized the DICOMLoader module from the open-source ROSAMLLIB library* to manage these relationships automatically.

To improve computational efficiency, all volumes were processed prior to segmentation. Each CT volume was cropped in the z-direction, cutting off superior and inferior slices, with these boundaries defined by the volume within 50% of the corresponding RTDose's maximum dose, expanded by a 2-slice buffer. A minimum axial length of 10 cm was enforced to ensure sufficient anatomical context for the deep-learning models. This same cropping strategy was applied to the RTDose volumes, ensuring that subsequent processing was restricted to the geometry containing the primary dose distribution.

The algorithm was developed and validated using two distinct datasets retrieved from our institution's Picture Archiving and Communication System (PACS). The training cohort consisted of 109 cases selected to intentionally balance the representation of each anatomic region. The testing cohort consisted of 100 consecutive treatment plans (excluding brachytherapy) extracted from the department's CT simulation schedule to represent a realistic clinical distribution. The distribution of anatomic region labels across both cohorts is detailed in Table 1. Note that the total number count of labels exceeds the number of patients, as complex plans often span multiple adjacent regions (e.g. "Thorax" plans leaking dose into the liver). All data collection and analysis procedures were conducted under a protocol approved by the UCLA Institutional Review Board (IRB-23-0371).

**Table 1**. Breakdown of the training and testing cohorts. Training data was selected to minimize class imbalance, while the testing cohort represents clinical prevalence based on 100 consecutive cases from the institution's CT simulation schedule.

|          | Cranial | H&N | Thorax | Abdomen | Pelvis | Extremity |
|----------|---------|-----|--------|---------|--------|-----------|
| Training | 20      | 17  | 27     | 20      | 29     | 20        |
| Testing  | 11      | 15  | 28     | 15      | 39     | 8         |

## 2.2. Autosegmentation

Three modules from Totalsegmentator[21-23] (total, head_glands_cavities, appendicular_bones) were used for autosegmentation of normal organs. These three modules contour a large majority of the body, covering many of the important organs treated in radiation oncology. The 'total' module was run in the 'fast' mode, contouring the organs in 3 mm isotropic resolution before returning to the original CT space. During parsing, the small bowel, colon, and duodenum masks from the 'total' model were combined and split into a pelvic bowel and abdominal bowel. In most cases, organs that span across multiple regions (e.g., esophagus) were deliberately ignored, but the bowels were deemed important enough in radiation oncology to split. The superior aspect of the hip bones was selected as the anatomical landmark for this division, as it is consistently and accurately segmented by Totalsegmentator, providing a reliable and verifiable boundary for separating the pelvic and abdominal bowel sections.

---

* https://github.com/YAAbdulkadir/rosamllib

If non-empty rib contours are detected, an in-house breast auto-contouring module is triggered. The MedNeXt[24] architecture was adopted to improve computational efficiency compared to the standard TotalSegmentator breast module. The model processes Z-normalized CT patches to predict left and right breast masks. Inference utilizes a sliding-window approach with consensus voting to ensure spatial continuity, followed by connected-component filtering to remove small artifacts.

### 2.3. Dose-mask Intersection Analysis

After generating the OAR masks, the workflow quantifies the target dose overlap. First, each RTDose volume is resampled to match the image grid (i.e., spacing, orientation, origin) of its corresponding cropped CT volume. A shape verification step ensures perfect spatial alignment between the dose and the deep-learning masks.

As a final optimization, a tighter bounding box is created using the precise extent of the volume within 50% of the maximum dose. This "isodose box" is applied to both the RTDose and all corresponding OAR masks. This step minimizes the data size by discarding empty voxels, enabling the volumes to fit into GPU memory for accelerated processing. While this additional cropping introduces a computational overhead, the efficiency of parallelism yields a net reduction in processing time.

### 2.4. Automated Anatomic Region Labeling Algorithm

For each plan, the algorithm identifies all organs with significant dose overlapping. This is determined by calculating a 'dose-overlap score' for each organ (*O*) against both the volumes within 85% and 50% of the maximum dose (*D*). This score is defined as:

$$Score = \frac{V(O \cap D)}{V(D)} \quad (1)$$

where *V(X)* represents the voxel volume of structure or intersection **X**, and the ratio therefore represents the fraction of the isodose volume occupied by organ **O**.

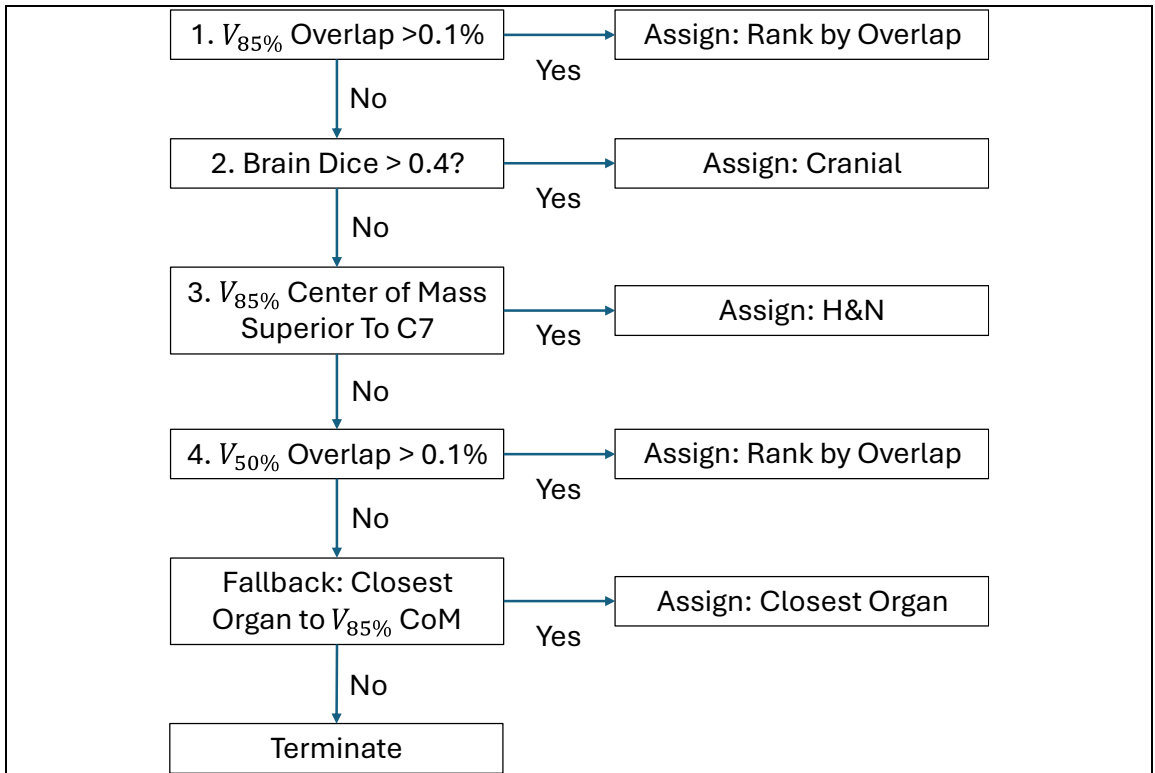

**Figure 1.** Hierarchical decision tree for anatomic region labeling. The algorithm prioritizes physical overlap with the high-dose region ($V_{85\%}$), utilizing geometric constraints and lower-dose thresholds ($V_{50\%}$) as sequential fallbacks for plans where the target coverage is ambiguous or small.

All organs with a score greater than 0.001 were first identified. This threshold was chosen empirically from the 109 curated development cases to exclude <0.1% overlaps that correspond to numerical noise. Each organ was mapped to an anatomical region using a predefined dictionary (see **Supplementary Table 1**). Organs that belong to the "Excluded" category, such as vessels that span multiple regions, were excluded to prevent ambiguous or misleading regional labels.

The prediction followed the hierarchical algorithm below (visualized in **Figure 1**) to label each treatment plan:

1. **Primary High-Dose Score ($V_{85\%}$)**: Identify all organs overlapping the volume within 85% of the maximum dose. Anatomical regions are ranked by their highest single-organ overlap score with each region included only once.
2. **Special Conditions**: If no organ has a Score > 0.85, apply the following checks:
   - **Cranial**: If the Brain Dice Score Coefficient is > 0.4 (empirically derived to exclude dose leakage), label as cranial.
   - **Head and Neck**: If the $V_{85\%}$ Center of Mass (CoM) is superior to the C7 vertebrae (or the Thyroid, if C7 is absent), label as Head and Neck.

3. **Secondary Moderate-Dose Score ($V_{50\%}$)**: If the plan remains unlabeled, repeat the overlap analysis using the volume within 50% of the maximum dose.
4. **Proximity Fallback**: If no organ has a Score > 0.50, assign the label of the organ closest axially to the original $V_{85\%}$ CoM.
5. **Termination**: If no proximity match is found, return no label to reduce false positives.

## 2.5. Implementation and Output

For each treatment plan, the workflow outputs:

1. A JSON file containing per-organ overlap and Dice scores for the volumes within 50% and 85% of the maximum dose.
2. The automated anatomic region label (e.g., Thorax, Abdomen) based on the organs that overlapped above a given threshold.

Together, these outputs form a scalable and interpretable system for labeling radiotherapy plans based purely on geometric dose-anatomy relationships, requiring no text-based plan name or target structure name information. This approach enables robust, high throughput labeling suitable for large institutional datasets where naming conventions are inconsistent or absent.

The complete workflow (segmentation → dose overlap → label assignment) was deployed on a workstation equipped with an NVIDIA 1080 TI GPU. The average processing time per plan was **133 seconds**, with the deep-learning segmentation accounting for the majority of runtime. While this represents a non-trivial computational requirement, it is a one-time preprocessing cost that generates permanent, reusable metadata for the database. Because the system's throughput is strictly bound by segmentation inference times, the overall speed of the curation pipeline scales directly with GPU hardware capabilities.

## 2.6. Validation Study

To evaluate the workflow, we established an expert labeling protocol to serve as ground truth and defined specific metrics for accuracy.

### 2.6.1. Expert-Based Anatomic Region Classifier

Accuracy of the site labeling algorithm was determined versus an expert-driven manual labeling protocol which was generated as follows.

1. An anatomic region was labeled by an expert if the 85% isodose volume penetrated a corresponding organ belonging to that region to a depth of 5 mm or more. Organs in the "Excluded" section of our anatomic region dictionary (see Supplementary Table S.1) were excluded from the labeling as they span multiple regions. Femoral heads were likewise excluded.
2. If multiple regions were identified from Rule 1, the order of regions was determined as follows:
    a. First, the 2D plane (coronal or sagittal) that displayed the largest cross-sectional area of the volume within 85% of the maximum dose was visually inspected for measurement.

b. Within this selected plane, the greatest linear dimension of the volume within 85% of the maximum dose, as it intersected with organs of each specific region, was measured using MIM Software electronic ruler tool.
   c. The anatomic regions were then ordered based on this measurement, from the longest dimension to the shortest.
3. If no organs met the 5 mm penetration criteria from Rule 1, the organ axially closest to the maximum dose point was identified, and its anatomic region was assigned as the region label.
4. Bowel contours were counted as 'Pelvis' if located below the superior aspect of the pelvic bone (iliac crest), and as 'Abdomen' otherwise.
5. If the volume within 85% of the maximum dose was completely above the C7 vertebrae and did not meet criteria for any other anatomic regions, it was considered 'Head and Neck'.
6. Plans targeting the entire brain according to visual interpretation were classified as 'Cranial' only (i.e., they were not assigned a secondary label of Head and Neck).

The above rules were optimized to match the intuition of expert radiotherapy planners when labeling plans while avoiding the subjectivity and irreproducibility of an intuitive expert labeling. In rule 1, this penetration depth of 5 mm was manually verified using the electronic ruler tool in MIM Software v7.4.3 (MIM Software Inc., Cleveland, OH), measuring orthogonally from the organ's surface inward to the isodose line. This threshold of 5 mm was selected to distinguish true anatomic involvement from standard inter-observer contouring variability, as the 95th percentile Hausdorff Distance ($HD_{95}$) up to 5 mm are frequently observed between expert benchmarks[25,26].

The femoral heads were excluded as AAPM Task Group 263 Report[16] implicitly treats them as pelvic OARs by including them in standard pelvic contouring templates (e.g., Gynecologic, Prostate, Anal cancer templates). Additionally, common clinical practice (and CPT coding) often treats the hip joint/femoral head as a distinct region or part of the pelvic girdle rather than a free extremity.

### 2.6.2. Evaluation Metrics

To evaluate the algorithm, a set of metrics were chosen to reflect different potential clinical use cases. For example, a user may need to filter for plans where "Thorax" is the primary label, while another user may need to find all plans that involve the thorax, even secondarily.

First, we measured three top-level accuracy scores to assess overall correctness of the labels. The most stringent, **Exact Accuracy**, counts a plan as correctly labeled only if the automated labels are identical to, and in the same order of importance as, the expert labels (e.g., a prediction of "Abdomen, Pelvis" would only match an expert label of "Abdomen, Pelvis"). The second, **Top-2 Accuracy**, is less strict, counting a plan as correct if the set of the top two automated anatomic region labels matches the set of the top two expert labels, regardless of order (e.g., "Abdomen, Pelvis" would be a match for "Pelvis, Abdomen"). The third, **Top-1 Accuracy**, focuses only on the most important label, counting a plan as correct only if the algorithm's top-ranked label matches the expert's top-ranked label, regardless of any other secondary labels.

To further analyze failure modes, we calculated the Sample-Based Precision and Recall (evaluation performance per-plan) and Label-Based Precision and Recall (evaluating performance per-anatomic region, e.g., reliability of the "Thorax" label specifically).

## 3. Results

The automated labeling algorithm was evaluated against the expert-labeled dataset of 100 consecutive treatment plans. The performance was assessed using plan-level accuracy metrics as well as per-class precision and recall.

### 3.1. Plan-Level Accuracy

We first evaluated the algorithm's overall performance on a per-plan basis. The algorithm achieved an Exact Accuracy of 91.0% (95% CI: 85.4-96.6%), meaning 91 out of 100 plans had automated labels that perfectly matched the expert labels in both content and order. The Top-2 Accuracy was 94.0% (95% CI: 89.3-98.7%), meaning the correct set of anatomic regions for 94 out of 100 plans were automatically labeled within the top two expert labels. Furthermore, the algorithm achieved a Top-1 Accuracy of 95.0% (95% CI: 90.7-99.3%), correctly identifying the region that received the highest percentage of treatment dose. These results demonstrate strong agreement with manual labeling, exceeding 90% for all metrics. The few observed discrepancies were largely confined to borderline, ambiguous cases.

### 3.2. Per-Class Performance

To understand the algorithm's performance for each specific anatomic region, we calculated the label-based True Positive, False Positive, True Negative, and False Negative counts, along with precision, recall (sensitivity), F1-score, and Top-1 Accuracy for each class. The results are summarized in **Table 2**.

The algorithm demonstrated perfect (1.0) Top-2 sensitivity for Abdomen, Extremity, Thorax, Cranial, and Head and Neck, indicating that each label the expert found, no matter the rank, was also found by the automated algorithm. It also achieved perfect (1.0) Top-2 precision for Cranial and Head and Neck, meaning it never assigned those labels incorrectly (0 False Positives). Consequently, as the F1-score balances precision and sensitivity, it was also perfect (1.0) for these two anatomic regions.

While the overall Top-2 sensitivity was high, the per-class Top-1 Accuracy (Table 2) reveals a more nuanced performance, with the algorithm being least accurate at identifying the primary label for Thorax (0.97). The lowest performing metric overall was the Top-2 precision for the Extremity class (0.80). As our predictive and expert algorithms label anatomic regions in order of maximal dose coverage within specified isodose lines, this value indicates the agreement in what should be considered the most important region.

**Table 2**. Displays the True Positive (TP), False Positive (FP), True Negative (TN), and False Negative (FN) values for each anatomic region across all 100 plans in our dataset. The Top-1 Accuracy determines the percentage of correctly automated primary anatomic region labels compared to our expert labels.

|  | Abdomen | Pelvis | Extremity | Thorax | Cranial | Head & Neck |
|---|---|---|---|---|---|---|
| TP | 15 | 38 | 8 | 27 | 11 | 16 |
| FP | 1 | 1 | 2 | 2 | 0 | 0 |
| TN | 84 | 60 | 90 | 71 | 89 | 84 |
| FN | 0 | 1 | 0 | 0 | 0 | 0 |
| Top-1 Accuracy | 0.98 | 0.98 | 0.99 | 0.97 | 1 | 0.98 |
| Top-2 Accuracy | 0.99 | 0.98 | 0.98 | 0.98 | 1 | 1 |
| Sensitivity (Top-2) | 1 | 0.9744 | 1 | 1 | 1 | 1 |
| Precision (Top-2) | 0.9375 | 0.9744 | 0.8 | 0.9310 | 1 | 1 |
| F1 Score (Top-2) | 0.9677 | 0.9744 | 0.8889 | 0.9643 | 1 | 1 |

## 4. Discussion

The results of our workflow on 100 consecutive treatment plans provide a strong indication of how it will work in a clinical setting. Running this dataset through the workflow resulted in an exact accuracy of 91.0%, a Top-2 accuracy of 94.0%, and a Top-1 Accuracy of 95.0%.

A key finding is that the algorithm never produced a non-adjacent region error. An example of this type of error would be the mislabeling of an Abdomen plan as "Abdomen, Head and Neck" skipping over the "Thorax." In other words, when multiple regions are labeled, they are contiguous, which is clinically and anatomically sound. The primary source of differences between the automated and expert labels was instead found in minor, adjacent-region discrepancies, most of which stemmed from logical differences between the automated algorithm and the manual protocol.

From a clinical operations perspective, this labeling system could serve as an automated indexing tool within institutional data warehouses, accelerating cohort identification for AI model training or retrospective dose-response analyses.

### 4.1. Analysis of Mismatched Cases

A review of the mismatched cases revealed that, with the exception of one systematic algorithmic failure, discrepancies predominantly occurred in "borderline" clinical scenarios where the ground truth was ambiguous, forcing the expert and the algorithm to prioritize different interpretations. These discrepancies generally fell into three categories: sensitivity to grazing dose, segmentation granularity constraints, and contouring variations.

The systematic failure occurred when the target dose did not clearly overlap any autosegmented organ, forcing a reliance on "tie-breaker" logic. In these spatially ambiguous cases, the expert protocol defaults to the organ closest to the point of maximum dose on a single axial slice. In contrast, the automated algorithm finds the organ closest axially to the calculated Center of Mass (CoM) of the dose within 85% of the maximum dose. **Figure 2** demonstrates a representative "Pelvis vs Extremity" case where the target dose was situated between the pelvis and the femur. While the expert could visually identify that the dose targeted the anus (a pelvic structure), the automated system lacked this context. Because the upstream TotalSegmentator model does not currently segment the anus[21], the algorithm detected no overlap and reverted to the fallback logic. Since the CoM of the dose fell inferiorly, it was geometrically closer to the femurs than the pelvic bones, triggering the "Extremity" label. This misclassification occurred not because the dose missed the anatomy, but because the targeted organ was missing from the model. This highlights a systematic failure that is likely to persist until more granular segmentation models are integrated.

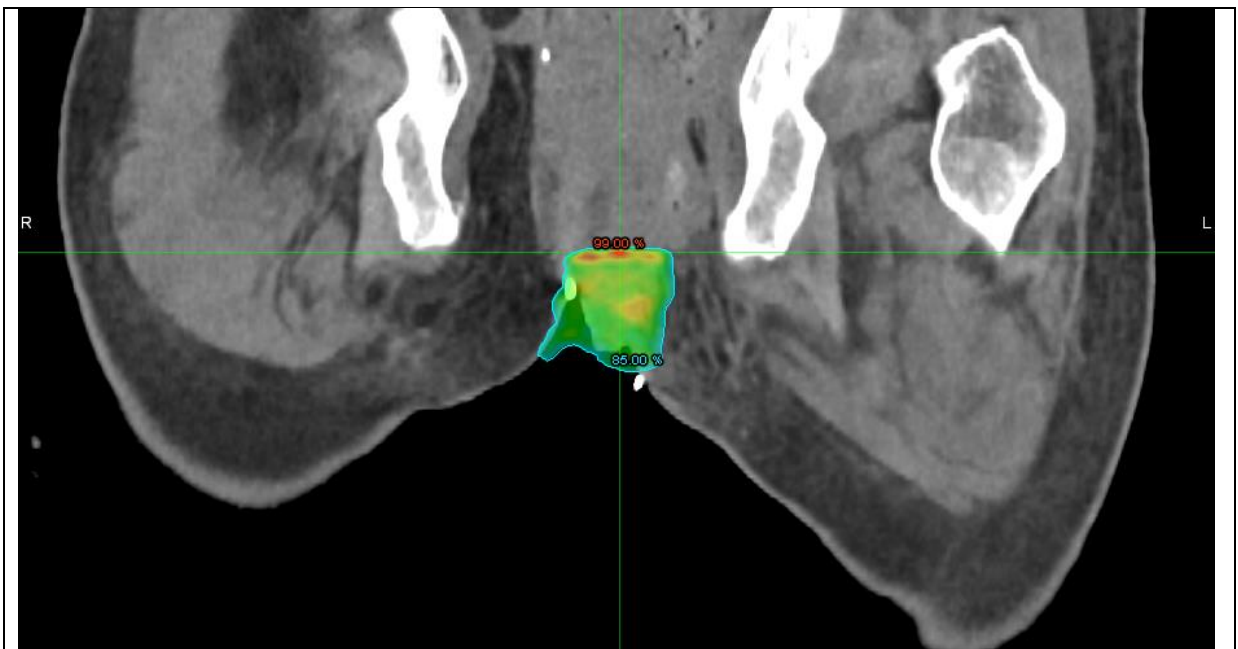

**Figure 2**. Divergent classification in a borderline Pelvis/Extremity case. The expert labeled the plan "Pelvis" based on the visual identification of the target. While the predictive algorithm assigned "Extremity" because the volumetric Center of Mass of the $V_{85\%}$ isodose (green) fell inferiorly to the pelvic bone, causing the closest structure to be the femur. Notably, in this specific instance, the experts fallback method of maximum dose (red) would have successfully captured the "Pelvis" label.

The first category involved threshold sensitivity, where the dose "grazed" the boundary of an anatomic region. The expert protocol required a strict 5 mm penetration depth to assign a label, while the automated algorithm was designed to be more sensitive, detecting overlap greater than 0.1%. **Figure 3** illustrates a plan where the target volume grazed organs in both the "Head and Neck" and "Thorax" regions but did not penetrate any of these organs by more than 5 mm. So, using the fallback method the expert labeled this plan "Head and Neck" as the fallback method forces a single closest anatomic region. On the other hand, the predictive algorithm detected dose in the thyroid, left clavicle, and left upper lung and assigned a label of "Thorax, Head and Neck".

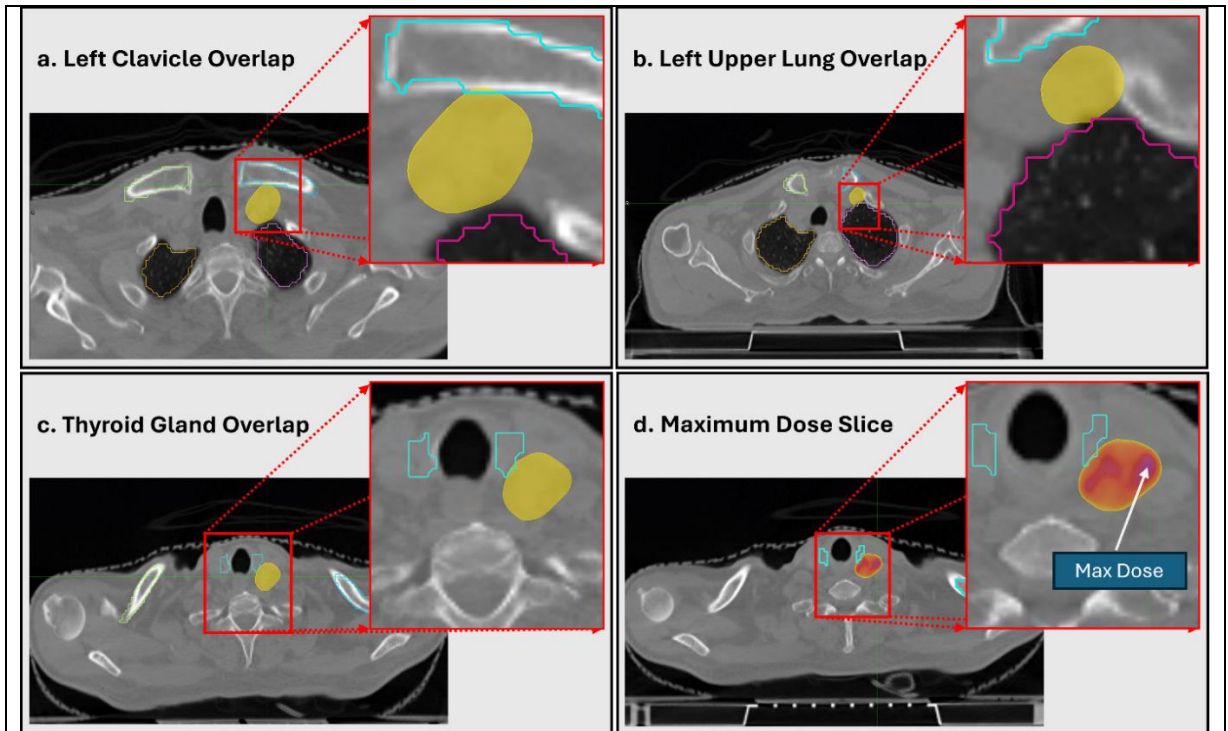

**Figure 3**. Displays the slice with the largest overlap in each organ (a, b, c), as well as the slice with the maximum dose (d). The point of maximum dose in (d) is shown by the red arrow. The expert label was determined from the closest organ to the point of maximum dose, axially, as no organ was penetrated to a depth of 5 mm. This led to an expert label as "Head and Neck." The predictive algorithm labeled the plan as both "Thorax" and "Head and Neck" as the overlap of the dose on the segmented organs was higher than the set threshold.

The remaining discrepancies stemmed from differences in the upstream autosegmentation inputs rather than the labeling logic itself. One "Extremity" false positive was caused by the algorithm's inability to distinguish the femoral head from the femoral shaft. In contrast, the expert protocol explicitly excluded the femoral head, consistent with the standard pelvic definitions[16], to avoid mislabeling pelvic plans. Finally, a single "Abdomen" false positive was caused by a variation in contouring styles, where the deep-learning model segmented the liver wrapping further around the inferior vena cava than the human expert, creating a dose overlap that did not exist in the human-drawn structures. These examples suggest that many "false positives" in our validation were instances where the automated workflow captured valid overlaps that the rigid expert protocol was forced to ignore.

### 4.2. Interpretation of Mismatches and Methodology

It is important to contextualize the Top-1 Accuracy "failures" described in Section 4.1, as this metric represents the primary utility for clinical database querying. In instances where the automated label did not match the expert label, such as the Head and Neck plan automatically labeled both Head and Neck and Thorax, the automated output remained clinically coherent[27]. While these predictions technically diverged from the rigid expert protocol, from the perspective of a radiation oncologist or physicist searching a database, these "mismatched" plans would likely still be relevant to a query for the automatically labeled anatomic region.

A primary challenge in establishing the validation framework was defining a robust, reproducible threshold for expert classification. A geometric constraint of 5 mm penetration was selected for a manual protocol to minimize inter-observer variability[7] and prevent the over-labeling of "grazing" fields, similar to the exclusion of the femoral heads. However, this geometric rule is merely an approximation of whether an organ is involved.

In contrast, the automated algorithm utilizes a volumetric dose-based metric, specifically the percentage of isodose volume overlaps. This approach is arguably more representative of the physical treatment delivery[28], as it accounts for the magnitude of the target volume; a large target requires more substantial organ overlap to trigger a label than a small, highly conformal target. Consequently, many of the "false positives" occurred when the algorithm detected overlaps that the expert protocol excluded. These cases likely represent genuine dosimetric overlap that simply fell below the strict geometric thresholds of the expert protocol.

### 4.3. Limitations

While the proposed workflow demonstrates high accuracy, it is subject to several key limitations. First, the algorithm's performance is fundamentally dependent on the fidelity of the upstream TotalSegmentator models. As observed in the analysis of discrepancies, segmentation errors propagate directly to the final classification. This was evident in the false-positive "Extremity" predictions driven by the inability to distinctively segment the femoral head nor segment the anus, as well as the "Abdomen" false positive caused by minor contouring variances in the inferior vena cava. Without a mechanism to correct these upstream segmentation inaccuracies, the dose-overlap logic will remain vulnerable to artifacts in the input structures.

Furthermore, this validation was restricted to a single-institution cohort of 100 consecutive plans. Although this dataset reflects a realistic local clinical mix, it does not capture the full extent of inter-institutional heterogeneity in planning styles or patient anatomy. Consequently, the heuristic thresholds empirically derived during development, specifically the 0.4 Dice score for cranial classification and the C7 vertebral boundary for Head and Neck stratification, require validation on external datasets to ensure robust generalizability. Finally, the expert labeling protocol itself, while designed for objectivity, prioritized consistency over the nuance of rare clinical edge cases.

### 4.4. Future Work

Future development will prioritize addressing these identified limitations to facilitate the integration of this algorithm into automated quality assurance and data curation pipelines. To mitigate "Extremity" false positives, a geometric post-processing step will be implemented to computationally separate the femoral head from the femoral shaft, thereby enabling the correct application of the femoral head exclusion rule. To further resolve the 'missing organ' failure mode identified in pelvic targets, future iterations will incorporate more granular segmentation models (e.g., anal canal, mediastinum).

Additionally, the "closest organ" fallback logic will be refined to improve consistency in ambiguous cases. We intend to transition from a purely geometric CoM, which weighs all voxels within the isodose volume equally, to a dose-weighted Center of Mass (wCoM) calculation, defined as:

$$R_{wCoM} = \frac{\sum(D_i \cdot r_i)}{\sum D_i} \qquad (2)$$

where $D_i$ is the dose at voxel $i$ and $r_i$ is the position vector. By weighting the spatial coordinates by the local dose, the fallback will more accurately reflect the high-dose core of the treatment, reducing sensitivity to asymmetric dose "leakage" that currently triggers incorrect adjacent-region labels. Crucially, the workflow requires validation on larger, multi-institutional datasets to confirm its generalizability and clinical readiness beyond a single center.

A critical long-term objective is the derivation of granular, treatment-site-specific labels. For instance, classifying a plan not merely as "Pelvis," but more descriptively as "Prostate" or "Prostate + Lymph Nodes." However, achieving this level of specificity presents a significant technical challenge, as it necessitates higher-fidelity segmentation than is currently provided by standard OAR models. Since incorporating additional specialized segmentation models inevitably increases computational overhead, the trade-off between granular clinical detail and processing efficiency remains a key area of investigation.

## 5. Conclusion

Effective data curation is a fundamental prerequisite for large-scale research, yet it is frequently impeded by the unreliability of heterogenous treatment plan nomenclature. To address this, we developed and validated a fully automated labeling framework that operates independently of text metadata, relying instead on objective dose-volume metrics and deep-learning autosegmentation.

In a validation cohort of 100 consecutive treatment plans, the algorithm demonstrated robust performance, achieving 91% Exact Accuracy, 94% Top-2 Accuracy, and 95% Top-1 Accuracy. The high Top-1 Accuracy is particularly clinically significant, indicating that the system can reliably facilitate cohort retrieval based on the primary treatment site. Although the workflow remains sensitive to upstream segmentation fidelity, it offers a scalable, objective, and standardized solution for the curation of multi-institutional datasets essential for the advancement of 'big data' applications in radiation oncology.

**Supplementary**

| Anatomic Region | Structures Included |
|---|---|
| **Cranial** | brain; optic nerve (left, right) |
| **Head and Neck** | auditory canal (left, right); hypopharynx; nasal cavity (left, right); parotid gland (left, right); submandibular gland (left, right); hard palate; soft palate; thyroid gland; eye (left, right); eye lens (left, right); nasopharynx; oropharynx; vertebrae C1–C7 |
| **Thorax** | ribs (left/right 1–12); vertebrae T1–T12; sternum; scapula (left, right); heart; left atrial appendage; costal cartilages; lungs (upper/lower/middle lobes, left/right); clavicle (left, right); breasts (left, right); nipples (left, right) |
| **Abdomen** | liver; pancreas; spleen; stomach; gallbladder; adrenal gland (left, right); kidney (left, right); vertebrae L1–L4; abdominal bowel |
| **Pelvis** | urinary bladder; prostate; hip (left, right); sacrum; vertebrae L5, S1; pelvic bowel |
| **Extremity** | femur (left, right); tibia; fibula; radius; ulna; humerus (left, right); patella; carpal; metacarpal; metatarsal; tarsal; phalanges (hand, feet) |
| **Excluded** | spinal cord; trachea; common carotid artery (left, right); autochthon (left, right); gluteus maximus/medius/minimus (left, right); iliopsoas (left, right); aorta; inferior vena cava; portal vein and splenic vein; pulmonary vein; superior vena cava; brachiocephalic trunk; brachiocephalic vein (left, right); subclavian artery (left, right); iliac artery (left, right); iliac vein (left, right); skull; esophagus; colon; small bowel; duodenum |

**Table S.1.** The Anatomic Region column shows the anatomic region that each of the Structures Included organs are a part of. As a part of our workflow, the colon, duodenum, and small bowel are split into a pelvic bowel and abdominal bowel with the separation at the superior aspect of the hip bone. The Excluded section is a list of organs that are not included in the expert or predictive algorithm. Each organ listed is an organ that was contoured during our workflow.